\documentstyle[preprint,aps,epsf]{revtex}
\tightenlines
\def\Journal#1#2#3#4{{#1} {\bf #2}, #3 (#4)}

\def\NPB{{\em Nucl. Phys.} B}
\def\JPA{{\em Jour. of Phys.} A}
\def\PLA{{\em Phys. Lett.} A}
\def\PLB{{\em Phys. Lett.} B}

\def\PR{{\em Phys. Rev.}}
\def\PRB{{\em Phys. Rev.} B}
\def\PRD{{\em Phys. Rev.} D}

\def\RMP{{\em Rev. Mod. Phys.}}
\def\RPP{{\em Rep. Prog. Phys.}}
\def\PREPC{{\em Phys. Rep.} C}
\def\RNC{{\em Riv. Nuovo Cimento}}
\def\ZETF{{\em Zh. Eksp. Teor. Fiz.}}
\def\PRS{{\em Proc. Roy. Soc.}}
\newcommand{\be}{\begin{equation}}
\newcommand{\ee}{\end{equation}}
\newcommand{\bea}{\begin{eqnarray}}
\newcommand{\eea}{\end{eqnarray}}

\newcommand{\cK} {{\cal K}}
\newcommand{\mod}{({\rm mod}2)}
\newcommand{\hf} {{1\over2}}
\newcommand{\nonu}{\nonumber\\}
\newcommand{\cm} {{\cal M}^2_L}
\newcommand{\vphi} {\varphi}

\newcommand{\cL} {{\cal L}}

\newcommand{\ml}{m_L^2}
\newcommand{\tml}{\tilde m_L^2}

\begin{document}
\title{The antiferromagnetic $\phi^4$ Model, I. The Mean-field Solution}

\author{Vincenzo Branchina\thanks{branchina@crnvax.in2p3.fr}}
\address{Laboratory of Theoretical Physics, Louis Pasteur University\\
3 rue de l'Universit\'e 67087 Strasbourg, Cedex, France}

\author{Herv\`e Mohrbach\thanks{mohrbach@crnvax.in2p3.fr}}
\address{Laboratory of Theoretical Physics, Louis Pasteur University\\
3 rue de l'Universit\'e 67087 Strasbourg, Cedex, France\\
and\\
LPLI-Institut de Physique, F-57070 Metz, France}

\author{Janos Polonyi\thanks{polonyi@fresnel.u-strasbg.fr}}
\address{Laboratory of Theoretical Physics, Louis Pasteur University\\
3 rue de l'Universit\'e 67087 Strasbourg, Cedex, France\\
and\\
Department of Atomic Physics, L. E\"otv\"os University\\
Puskin u. 5-7 1088 Budapest, Hungary}
\date{\today}
\maketitle
\begin{abstract}
Certain higher dimensional operators of the lagrangian may render the vacuum
inhomogeneous. A rather rich phase structure of the $\phi^4$ scalar model in 
four dimensions is presented by means of the mean-field approximation. One finds 
para- ferro- ferri- and antiferromagnetic phases and commensurate-incommensurate transitions. 
There are several particles described by the same quantum field in a manner similar
to the species doubling of the lattice fermions. It is pointed out that chiral bosons
can be introduced in the lattice regularized theory.
\end{abstract}

\section{Introduction}
Only renormalizable Quantum Field Theory models are considered in 
Particle Physics. This was explained traditionally by 
inspecting the UV divergences generated by 
the operators in the framework of the perturbation expansion 
in a homogeneous background field. The non-renormalizable
theories were rejected due to the need of infinitely many coupling constants. 
This argument has been further developed in the last 
decades. First it came the realization that what really matters 
in Particle Physics are not the true UV divergences because one 
always works with effective theories in the lacking of definite knowledge 
of the Theory of Everything. The characterization of the renormalizable
operators was modified by looking into their importance at
low energies. In particular the equivalence of the renormalizability
of an operator with its relevance at the UV fixed point
has been established \cite{wilsrg}. Non-renormalizable operators are
excluded because they do not change the universality class, i.e. their
influence on the dynamics decreases as we move
away from the UV scaling regime towards the physical energy scales. 

There are different mechanisms which nevertheless may turn a coupling 
constant which is found irrelevant in the usual treatment into an important 
parameter of the theory.

{\em Loop corrections:} 
Once the anomalous dimension is taken into account 
in the power counting argument \cite{bard} 
new relevant operators at the UV fixed point can be generated. 
The physical picture 
of the strong coupling massless QED vacuum \cite{miransky} which
suggested this natural generalization of the power counting method is based 
on the observation that the positronium may acquire a negative energy and 
collapse onto the size of the cut-off for $e=O(1)$. 
The condensate of these bound states breaks the chiral symmetry and the
IR features of the resulting vacuum are modified when compared to the perturbative 
ones. 

{\em Multiple fixed points:}
Starting from the Theory of Everything a unique renormalized trajectory 
describes the Composite Models, Grand Unification, Electro-weak Theory, QCD,
QED, Condensed Matter Physics and Solid State Physics which appear as different
"scaling islands" in the coupling constants space. One operator which is 
irrelevant in the vicinity of one such fixed point may turn to be relevant around 
another one \cite{senben}.

{\em Tree-level effects:}
The power counting argument traces down the influence of the loop corrections
to the scaling laws. The tree level effects of certain operators might be
much more complicated in relating different length scales and they may 
generate new important coupling constants which defy the classification 
based on a perturbative implementation of the Wilson-Kadanoff blocking 
procedure \cite{enzoe}. 
The cut-off is usually ignored
in the tree-level solution though it is actually present in any consistent regularization
of the path integral. When the important configurations, the saddle points, 
have length scales close to the cut-off these tree-level cut-off
effects become more important \cite{enzok}. In fact, if the semi-classical 
vacuum is non-homogeneous the successive elimination of the
degrees of freedom in the blocking procedure should be performed in the
semi-classical approximation. The non-trivial saddle points generate 
new contributions to the scaling laws \cite{jean}.

The surprisingly strong tree-level effects of the non-homogeneous saddle 
points raise the following issue. The usefulness of the ferromagnetic 
condensate in mass generation has long been been recognized in Particle
Physics \cite{najo}. The vacuum is a coherent state of 
particles with zero momentum. What happens if particles with non-vanishing
momentum form a condensate ? A close similarity can 
be found in the charge or spin density wave phases of solids \cite{gruner}.
The emergence of these states from the normal ground state is a
highly involved dynamical non-equilibrium problem. Here we are
interested in the properties of the static modulated phase only.

The field expectation value in this vacuum 
is a non-trivial function of the space-time coordinates with a characteristic
length given by the inverse of the typical momentum of the particles in the
coherent state. 

An oscillating saddle point configuration is formally
similar to the N\`eel state of the antiferromagnetic Ising model,
the ground state of solids or the charge density wave state. 
Such a formal similarity leads to far reaching analogies between Solid State
Physics and the phenomenology of this condensate. The nonvanishing
momentum of the condensed particles extends the symmetry breaking from
the internal symmetries to the external ones. The result is the dynamical
breakdown of the space-time symmetries. It is well known that
the ground state of the translational invariant hamiltonian for
photons, electrons and massive positively charged ions is not
translational invariant for certain densities
(solid state crystals). Due to the observed homogeneity of the space-time
in the particle reactions the absence of the dynamical breakdown
of the external symmetries was always assumed \cite{ass}, or 
taken for granted in high energy physics. 

In this paper we will consider the case of a single component
self interacting scalar field theory in the presence of higher
derivative terms. The semiclassical
solution of our model reveals the 
possibility of breaking the space-time
symmetries at the cutoff scale in a manner 
which is compatible with the homogeneity 
of the space-time at finite observational scales. The spontaneous breakdown 
of the internal symmetries 
is widely accepted and used
in Quantum Field Theories. The saddle point
approximation which is based on the construction of a 
condensate in the vacuum can be used without difficulties in 
exploring the possibility of the spontaneous breakdown of
external symmetries, too. In this case the condensate is
obviously inhomogeneous. This inhomogeneity amounts
to a periodic structure in our case. The apparence 
of an elementary cell repeated periodically in the
vacuum manifests itself in the possibility of exchanging nonvanishing
momentum between the propagating particles and the particles 
condensed in the vacuum and in the presence of several branches in
the dispersion relation, such as the acoustic and optical phonons
in the solid state crystals. The vacuum of the model considered in this
paper consists of a condensate of particles with a given momentum,
$p^\mu=P^\mu$.
Thus the inhomogeneity of the vacuum leads to the nonconservation
of the momentum, $p^\mu\to p^\mu\pm P^\mu$. We constrain the 
external symmetry by allowing those translations only which bring 
one elementary cell into another. The momentum defined by this subgroup of the
original external symmetry group, $p_{phys}^\mu=p^\mu mod(P^\mu)$, 
is the analogous of the Bloch momentum in Solid State Physics and it is 
obviously conserved. The different branches of the
dispersion relation are interpreted as different "flavor" states
of the elementary excitations, the particles of the model. The 
umklapp process, where a
momentum $\pm P^\mu$ is exchanged with the vacuum is then
a "flavor" changing reaction. Thus the momentum nonconservation is
traded into a "flavor" nonconserving dynamics where the particles in
a given "flavor" state propagate in a homogeneous vacuum. When the
momentum $P^\mu$ diverges then the space-time structure of the
"flavor" changing processes remain unresolved for the low (physical) 
momentum observables.

In the lattice regularized version of our model
we find the usual phases of Solid State Physics which 
belong to the para- ferro- ferri- or antiferromagnetic vacuum. 
This rich phase structure is due to the presence of more than next 
neighbour couplings. In this case the dispersion relation (the quadratic 
part of the action at vanishing field) has two mimima 
at $p^\mu=0$ and at $p^\mu=P^\mu=\frac{\pi}{a}$. The ferromagnetic
or antiferromagnetic vacua occur when the first or 
the second minimum become negative. Having taken into account the condensation 
of the instable modes,
the two minima of the new dispersion relation correspond to elementary excitations
which allow us to identify two particles described by the same quantum field.

The unusual features
of the model can be traced back to the fact that the condensate and the dynamical
symmetry breaking are driven by the kinetic rather than the potential energy terms 
of the action. It has already been remarked that the kinetic energy becomes
dominant in high temperature QCD and leads to the dynamical breakdown of the 
fundamental group symmetry in high energy processes \cite{qcdht}. This time the
kinetic energy drives the formation of the non-trivial elementary cells in the ground state 
which break the space-time inversion symmetries and introduce a non-trivial length scale
in the vacuum.

Here we consider the theory
in the mean-field approximation. The organization of the paper is 
the following. In section 2 we introduce 
our higher derivative $\Phi^4$ scalar model. The tree level
phase structure of this model
in dimension $d\le4$ is presented in Section 3.
A more detailed analysis in the vicinity of the
simplest antiferromagnetic phase of $d=4$ is performed in Section 4.
The elementary excitations are identified by the help of the free propagator
in Section 5. The symmetry aspects of the phase diagram and the 
interpretation of the different particle modes of the system is the subject
of Section 6. Finally, Section 7 is for the conclusions.

\section{Higher derivative scalar model}
To study the impact of higher derivative terms in the case of the single component
scalar model we choose the following action
\be
S[\Phi(x)]=\int d^d x\biggl\{
\hf\partial_\mu\Phi(x)\cK\biggl({(2\pi)^2\over\Lambda^2}\Box\biggr)\partial_\mu\Phi(x)
+{m^2\over2}\Phi^2(x)+{\lambda\over4}\Phi^4(x)\biggr\},\label{lagrc}
\ee
where 
\be
\cK(z)=1+c_2z+c_4z^2.
\ee
The dimension of the higher derivative terms is taken into account by the 
introduction of the scale parameter $\Lambda$.

Models with higher order derivative terms have already been considered recently
\cite{zk}, \cite{parisi}, \cite{zh}, \cite{shamir}.
We suppose first that the vacuum is homogeneous, $<\Phi(x)>={\rm const}.$ 
The quantum fluctuations are then plane waves with the eigenvalues
\be
G^{-1}(p^2)=m^2+p^2-c_2(2\pi)^2{p^4\over\Lambda^2}+c_4(2\pi)^4{p^6\over\Lambda^4}
\label{invp}
\ee
for the second functional derivative of the action.
Modes with negative $G^{-1}(p^2)$ are unstable and generate a condensate.
When $m^2<0$ and $c_j=0$ we have a ferromagnetic instability.
After filling up the most unstable mode $p=0$, the system
stabilizes itself and the only change in the interaction is
the apparence of a three particle vertex which describes
the processes where a particle is exchanged with the condensate. If 
$G^{-1}(p^2)$
develops a second minimum at $p\not=0$, 
new particles appear in the system.
When $c_2$
becomes large (and for simplicity we limit ourselves to consider the case 
$m^2=0$) an instability shows up for 
\be
{c_2\over2c_4}-\sqrt{{c_2^2-4c_4\over4c_4^2}}<(2\pi)^2{p^2\over\Lambda^2}<
{c_2\over2c_4}+\sqrt{{c_2^2-4c_4\over4c_4^2}}.
\ee
Thus a condensate of particles with non-vanishing momenta 
is formed. The filling of this condensate by the most unstable mode,
$p^2_{cond}=c_2\Lambda^2/2(2\pi)^2c_4$, modifies the interaction between the plane wave 
modes in a rather complicated manner and particles with different momenta 
may appear in the stable condensate. It is reasonable to expect that the 
Fourier transform of the field expectation value is peaked around
$p^2=p^2_{cond}$. Its spread is a measure of the strength of the
interaction within the condensate. Notice that the action is bounded 
from below  for $\lambda>0$ because $S\to\infty$ either when $p^2\to\infty$ or 
when the amplitude of the oscillations tends to infinity.

We will explore the dynamics of the effective theory (\ref{lagrc})
in the mean-field approximation. The first step will be the determination of 
the phase structure. The spectrum of the free quantum fluctuations
will be studied later.

\section{The mean-field phase structure}
In this Section we classify the different phases of the scalar theory 
(\ref{lagrc}) at the tree-level. We identify the order parameter with
the condensate, $<\Phi(x)>$, which is a non-trivial function whose 
Fourier transform is
\be
\tilde\Phi_{vac}(p)=\int d^dxe^{ipx}<\Phi(x)>.
\ee
The ferromagnetic condensate is obtained for 
$\tilde\Phi_{vac}(p)=\Phi_0\delta^{(d)}(p)$. For an antiferromagnetic 
condensate the function $\tilde\Phi(p)$ has a peak at $p\approx p_{min}$. 
One may also have the ferrimagnetic phase
where $\tilde\Phi_{vac}(p)$ displays two peaks, one at $p=0$ and another
at $p=p_{min}\not=0$. When a consistent cut-off like the lattice regulator is 
used, there are two length scales in the antiferromagnetic vacuum, the 
periodic length of the condensate and the regulator itself. The phase structure is 
amazingly rich in this case due to the commensurate-incommensurate phase 
transitions. We begin our investigation by looking for the lowest action 
solution of the equation of motion for (\ref{lagrc}).

\subsection{d=1}
We have solved numerically the finite difference equation 
in two different
regimes, $\Lambda^{-1}>>a$ and $\Lambda^{-1}\approx a$, where
$a$ is the step size (regulator) of the equation.

\underline{The continuum theory, $\Lambda^{-1}>>a$:}
The removal of the cut-off $a$ is trivial for the tree level classical 
equation and one finds the solution of the differential equation as $a\to0$.
In order to find the lowest action solution we considered the configurations
which yield finite action density,
\bea
\epsilon&=&{1\over T}\int_0^Tdx\biggl\{\hf\partial\Phi(x)\biggl[1
+c_2{(2\pi)^2\over\Lambda^2}\Box+c_4{(2\pi)^4\over\Lambda^4}\Box^2\biggr]\partial\Phi(x)\nonu
&&+{m^2\over2}\Phi^2(x)+{\lambda\over4}\Phi^4(x)\biggr\}.\label{actden}
\eea
The solutions of the variational equation are not constant since 
$\epsilon(0)>\epsilon(p^2_{cond})$. At the same time the $\Phi^4$
potential energy keeps $|\Phi(x)|$ bounded. Thus we conjecture that all 
solutions with finite action 
density are periodic and we impose periodic boundary conditions 
on the field, $\Phi(x)=\Phi(x+T)$. The minimization of the action density 
with respect to $\Phi(x)$ and $T$ indeed leds to periodic solutions, one of 
them is depicted in Fig. 1a. It was checked by increasing the volume $T$
that the minimal action configuration remains periodic.

This result can be compared with the a variational approach where the form
\be
\Phi(x)=A\sin\omega x \label{ansa}
\ee
is assumed. The minimization of the action density with 
$T={2\pi\over\omega}$ yields
\be
\omega^2={c_2\Lambda^2\over3(2\pi)^2c_4}\biggl(1+\sqrt{1-\frac{3c_4}{c_2^2}}\biggr),
\ee
and
\be
\epsilon=-\frac{3\lambda A^4}{32}.
\ee

One can see that the periodicity length is determined by the $c_j$ and the
amplitude is controled by the term $\phi^4$. By choosing for example 
$c_2(2\pi)^2=1$, $c_4(2\pi)^4=0.1$, $m^2=-0.1\Lambda^2$ 
and $\lambda=0.1\Lambda^3$, these two methods give
$A_{num}=10.65\Lambda^{-1/2}$, $A_{var}=10.59\Lambda^{-1/2}$, 
$\epsilon_{num}=-120.75\Lambda$ and 
$\epsilon_{var}=-117.94\Lambda$.

\underline{Lattice regulated model, $\Lambda^{-1}\approx a$:}
The periodicity length, $\ell=O(p^{-1}_{cond})$, of the solution of the continuous differential 
equation is a  "floating", analytical expression of the coupling constants. 
This may change when finite difference equations are considered. In fact,
the periodicity length of the condensate may be incommensurate
\cite{commes} with the numerical discretization of the differential equation. 
$\ell$ and $a$ are commensurate and belong to the class $(M,N)$ 
when $M\ell=Na$, $M$ and $N$ being relative primes. 
In this case $\ell$ locks in as a function of the coupling constants and 
creates a devil's staircase. A similar phenomenon was analyzed for models 
where the kinetic energy is quadratic but has minimum at non-vanishing momenta
\cite{frenkont}. One can easily find the simple commensurate phases,
$M=1$, $N$ not too large, c.f. Fig. 2. The high $(M,N)$ commensurate
points are presumably washed together with the incommensurate regions when
the quantum fluctuations are taken into account. 

We will study the
particle content of the simplest antiferromagnetic phase, $(M,N)=(1,2)$
in the next Section. The generalization of our method for the higher 
commensurate theories is possible though complicated. The spectrum of the 
elementary excitations of the incommensurate theories is rather involved
and qualitatively different \cite{elexspic}.

The phases $(1,N)$ with odd $N$ are of ferrimagnetic type. This is because there are an odd
number of lattice field variables within a period  which in general do not
add up to zero, c.f. Fig. 1b.

The complex phase structure of the lattice theory should be present in
any other regularization as well when the regulator is introduced in a
consistent manner at the tree-level. One may take the sharp momentum space
cut-off, as an example. Its implementation on the tree-level leads
to the acceptance of field configurations as possible saddle points whose 
Fourier components are vanishing for momenta beyond the cut-off. We believe that
the solutions of the Euler-Lagrange equations which satisfy such
a constraint display similar commensurate-incommensurate transitions
though the details of the phase diagram may differ.

\subsection{d=2,3,4}
For $d>1$ the staggered order generated by $c_2$ is more
complicated. This is due to the fact that the kinetic energy of the continuum
theory is $O(d)$ invariant and the most unstable modes at the minimum of the
dispersion relation are found on a $S_{d-1}$ sphere. The minima of
the dispersion relation should form a discrete set of points in the 
restricted Brillouin zone in order to have particle like
excitations. The degenerate modes on this sphere may achieve this by
creating a complicated dynamical $O(d)$ symmetry breaking
pattern. The staggered
antiferromagnetic order can be realized in $d_{AF}$ dimensions, $0\le d_{AF}\le d$.
For the cases $d_{AF}=d,~d-1$ and $d_{AF}<d-1$ we will use respectively the names
relativistic, non-relativistic and anisotropic vacuum. 
 
\underline{Continuum theory, $\Lambda^{-1}>>a$:}
We found the local minima of the action density corresponding to the
relativistic and the non-relativistic vacuum in $d=2$ as depicted in Fig. 3.
The latter is the absolute minimum. Inspired by the numerical 
results we have tried the following ansatz,
\be
\Phi_{rel}(x,y)=A\sin\omega x~\sin\omega y,
\ee
and
\be
\Phi_{nrel}(x,y)=A\sin\omega x.
\ee
The variational method gives acceptable but less accurate result than in
the one dimensional case for $\Phi_{rel}(x,y)$ due to the tree-level 
interactions which split the degeneracy of the condensate at $|p|=p_{cond}$.
The highly nontrivial effect of such a deformation of the saddle
point on the elementary excitation will be investigated below.

The issue of the $O(d)$ symmetry breaking pattern can be better 
studied in lattice regularization where the
regulator is explicit already at the tree level. 

\underline{Lattice regulated model, $\Lambda^{-1}\approx a$:}
The lattice regulated action in $d>1$ dimensions written in terms
of the dimensionless variables $x^\mu$, $\vphi=a^{d/2-1}\Phi$, $m^2_L=m^2a^2$
and the unit vectors $(e_\mu)^\nu=\delta_{\mu\nu}$ is
\bea
S[\vphi(x)]&=&\sum_x\biggl\{-\hf\vphi(x)\biggl[
A\vphi(x)+\sum_\mu\biggl(B(\vphi(x+e_\mu)+\vphi(x-e_\mu))\nonu
&+&C(\vphi(x+2e_\mu)+\vphi(x-2e\mu))
+D(\vphi(x+3e_\mu)+\vphi(x-3e_\mu))\biggr)\nonu
&+&\sum_{\mu\ne \nu}\biggl(E(\vphi(x+e_\mu+e_\nu)+2\vphi(x+e_\mu-e_\nu)
+\vphi(x-e_\mu-e_\nu))\nonu
&+&F(\vphi(x+2e_\mu+e_\nu)+\vphi(x+2e_\mu-e_\nu)+\vphi(x-2e_\mu+e_\nu)\nonu
&+&\vphi(x-2e_\mu-e_\nu))\biggr)\nonu
&+&G\sum_{\mu\ne\nu\ne\rho}\biggl(\vphi(x+e_\mu+e_\nu+e_\rho)
+3\vphi(x+e_\mu+e_\nu-e_\rho)\nonu
&+&3\vphi(x+e_\mu-e_\nu-e_\rho)+\vphi(x-e_\mu-e_\nu-e_\rho)\biggr)\biggr]\nonu
&+&{m^2_L\over2}\vphi^2(x)+{\lambda\over4}\vphi^4(x)\biggr\}\label{res}
\eea
where the coefficients $A,B,C,D,E,F,G$ are defined by
\bea
A&=&-2d+(4d^2+2d)c_2-(8d^3+12d^2)c_4,\nonu
B&=&1-4dc_2+(12d^2+6d-3)c_4,\nonu
C&=&c_2-6dc_4,\nonu
D&=&c_4,\nonu
E&=&c_2-6dc_4,\nonu
F&=&3c_4,\nonu
G&=&c_4.\label{forgsz}
\eea

Only the ferromagnetic phase and the antiferromagnetic phase (1,2) were located 
(Fig.4) by a numerical minimization of the action.
The absolult minimum of the action is relativistic in the (1,2) antiferromagnetic
phase, the non-relativistic and anisotropic vacua lie higher as local minima.

\section{The $c_4=0$ phases}
We will determine the boundary of the para-, ferro- and the $(1,2)$ 
antiferromagnetic phases by means of the mean-field method.
In the rest of this paper we
will constrain ourselves to the case $c_4=0$.
The explicit apparence of the cut-off makes the action with $c_4=0$ 
bounded from below. 

We seek the vacuum in the form
\be
\vphi(x)=\vphi_0+\vphi_1(-1)^{\sum\limits_{\mu=1}^{d_{AF}}x^\mu},
\ee
where $\vphi_0$ and $\vphi_1$ are variational parameters and $d_{AF}$ is the 
number of antiferromagnetic directions. The action of the lattice Laplace operator 
on the vacuum is
\bea
\Box\vphi(x)&=&\sum\limits_{\mu=1}^{d_{AF}}
[\vphi(x+e_\mu)+\vphi(x-e_\mu)-2\vphi(x)]\nonu
&=&-4d_{AF}(\vphi(x)-\vphi_0),
\eea
which yields
\be
-\Box\cK(\Box)\vphi(x)=\cm(d_{AF},c_2)(\vphi(x)-\vphi_0),
\ee
where
\be
\cm(d_{AF},c_2)=4d_{AF}\cK(-4d_{AF})=4d_{AF}(1-4d_{AF}c_2).
\label{propg}
\ee
The minimization of the action density
\be
s(\vphi_0^2,\vphi_1^2)=\hf\ml\vphi_0^2+\hf(\ml+\cm)\vphi_1^2
+{\lambda\over4}(\vphi_0^4+6\vphi_0^2\vphi_1^2+\vphi_1^4),\label{varpo}
\ee
gives $d_{AF}=d$ and leads to the phase diagram in Fig.5.
\vskip 10pt
The $\ml<0$ case.
\vskip 10pt
The equation
\be
\cm=0 
\ee
is a ferromagnetic-antiferromagnetic transition line. 
Clearly for $c_4=0$ there is no frustration in the system 
because the coupling constants C and E are both positive and then both of the 
ferromagnetic type. For $\cm>0$ (i.e. $c_2<\frac{1}{4d}$) the B next to neighbour
coupling is also positive and then the phase is ferromagnetic. For the saddle point 
we find: $\vphi_0^2=-\frac{\ml}{\lambda},~~\vphi_1^2=0$. 
Nevertheless it is important to notice that this phase is very different from 
the standard ferromagnetic phase of the theory
without higher derivatives terms, where $C=E=0$ and $B>0$. In fact, as we will show 
later, in each phase of our model (the antiferromagnetic as well as the paramagnetic 
and the ferromagnetic ones) we find two kind of particles. 

On the contrary for $\cm<0$, 
B is negative (that is of the antiferromagnetic type) 
and the phase is antiferromagnetic. In this case the saddle point,  
is $\vphi_0^2=0,~~\vphi_1^2=-\frac{\ml+\cm}{\lambda}$. 

On the transition line $\cm=0$, we have $B=0$. The absence of interactions between
next neighbours causes the lattice to split into two different non interacting 
(even and odd) sublattices. Approaching this line from the
ferromagnetic side  we have a ferromagnetic condensate of the
same magnitude and sign in each of these sublattices. Approaching 
this line from the 
antiferromagnetic phase we have two ferromagnetic condensates 
of the same magnitude but opposite sign in the two sublattices. We will show later 
that on this line our theory is the superposition of two independent standard 
ferromagnetic $\phi^4$ models.
\vskip 10pt
The $\ml>0$ case.
\vskip 10pt
The transition line between the paramagnetic and the antiferromagnetic phases is 
given by the equation
\be
\ml+\cm=0. 
\ee
As befor at the line $\cm=0$, the next to neighbour coupling $B=0$. We will see 
later that on this line our model corresponds to the superposition of two standard
paramagnetic $\phi^4$ theories.
If $\cm>0$ then $B>0$ and the phase is paramagnetic as expected. 
For $-\ml<\cm<0$, $B$ is negative but the phase is still paramagnetic. The phase is antiferromagnetic
when $\cm<-\ml$.

\section{The elementary excitations}
The quasiparticles of the mean-field approximation are given by the help 
of the free propagator. We will obtain the propagator in the different 
phases considered above. We start with
\be
<\phi(x)\phi(y)>=\int_{|p|\le\pi}{d^dp\over(2\pi)^d}e^{-ipx}G(p),
\ee
where
\be
G^{-1}(p)=\tml+\hat p_\mu\hat p^\mu\cK(-\hat p_\mu\hat p^\mu),
\ee
and
\be
\hat p_\mu=2\sin{p_\mu\over2}.
\ee
The mass parameter is given by
\be
\tml=\cases{\ml&P,\cr-2\ml&F,\cr-2\ml-3\cm(d,c_2)&AF,}
\ee
in the different phases. We write
\be
G^{-1}(p)={\cal P}^2(p)-c_2{\cal P}^4(p)+\tml,\label{eq:G}
\ee
with the notation
\be
{\cal P}^2(p)=4\sum_\mu\sin^2{p^\mu\over2}.
\ee

The excitations may take a momentum
\be
P_\mu(\alpha)=\pi n_\mu(\alpha),
\ee
from the vacuum where $n_\mu(\alpha)=0$ or 1.
The relation between the index $1\le\alpha\le2^d$ and the vector 
$n_\mu(\alpha)$ is 
\be
\alpha=1+\sum_{\mu=1}^dn_\mu(\alpha)2^{\mu-1}.
\ee
It is then advantageous to split the 
\be
{\cal B}=\bigg\{k_\mu,~|k_\mu|\le\pi\biggr\}
\ee
Brillouin zone into $2^d$ restricted zones,
\be
{\cal B}_\alpha=\biggl\{|k_\mu-P_\mu(\alpha)|\le{\pi\over2}\biggr\}.
\ee

The fluctuations around an extremum which is at the same time a minimum
of the propagator are the particle like excitations. In this manner the
single quantum field $\phi(x)$ might describe several
particles at the same time. 
We will use the restricted zone notation in each phase and will see
that only the particle modes survive in the continuum limit.

\subsection{The extrema of the free propagator}
In order to distinguish the particle like modes from other excitations
we have to locate the extrema of the propagator. 
The derivative of the inverse propagator,
\be\label{eq:derGg}
{dG^{-1}\over dp_\sigma}=2\sin p_\sigma\biggl(1-2c_2{\cal P}^2(p)\biggr),
\ee
shows that the propagator has indeed $2^d$ extrema at the centers of the 
restricted Brillouin zones. The other extrema satisfy the equation
\be
{\cal P}^2(p)=\frac{1}{2c_2}.
\label{secmini}
\ee
which are maxima.
The inverse propagator takes the values 
\be
G^{-1}(P_\mu(\alpha))=\cm(d(\alpha),c_2)+\tml,
\ee
at the center of the Brillouin restricted zones where the
two variables function $\cm$ is defined in (\ref{propg}) and
\be
d(\alpha)=\sum_\mu n_\mu(\alpha)\label{ncomps}
\ee
is the number of dimensions with antiferromagnetic excitations.

The second derivative of the propagator is
\bea\label{eq:derG2}
\frac{\partial^2G^{-1}}{\partial p_\sigma\partial p_\rho}
&=&2\delta_{\rho\sigma}\cos p_\sigma(1-8c_2{\cal P}^2(p))\nonu
&+&8\sin p_\sigma\sin p_\rho(12c_4{\cal P}^2(p)-c_2).
\eea

\underline{The Brillouin zone ${\cal B}_1$:} 
We find
\be
\frac{\partial^2G^{-1}}{\partial p_i^2}\Big \vert_{p=P(1)}=2, 
\ee
so the Bloch waves of the longest wavelength zone are always particle like. 

\underline{The Brillouin zone ${\cal B}_{2^d}$:} 
\be
\frac{\partial^2G^{-1}}{\partial p_i^2}\Big\vert_{p=P(2^d)}
=-2(1-8dc_2).\label{fift}
\ee
The right hand side is positive, and ${\cal B}_{2^d}$
describes particle like excitations in the region of the coupling constant
space considered in the previous Section.

\underline{The Brillouin zones ${\cal B}_\alpha,~~\alpha=2,\cdots,15:$} 
We present here the case $\alpha=2$ only where $P_\mu(2)=(\pi,0,0,0)$,
\bea
\frac{\partial^2G^{-1}}{\partial p_1^2}\Big\vert_{p=P(2)}&=&-2(1-8c_2+64c_4),\nonu
\frac{\partial^2G^{-1}}{\partial p_\ell^2}\Big\vert_{p=P(2)}
&=&2(1-8c_2+48c_4),
\eea
for $\ell=2,3,4$ and
\be
\frac{\partial^2G^{-1}}{\partial p_\mu\partial p_\nu}\Big\vert_{p=P(2)}=0
\ee
for $\mu\not=\nu$. The other zones yield similar result and
they contain no extrema but saddle points only.

Thus one finds two particle modes in the phases considered.
The other 14 reduced Brillouin zones have excitations which are
non-particle type.

\subsection{The continuum limit}
In order remove the 14 unusual excitations found above we take 
the continuum limit, $a\to0$. This is quite a simple procedure in
the mean-field approximation where the quantum fluctuations are kept 
non-interacting. We keep the mass parameter $m^2$ of the lagrangian
cut-off independent in this approximation so $\ml=O(a^2)$.

The propagator 
\be
G^{-1}(p)=\tml+{\cal P}^2(p)-c_2{\cal P}^4(p),
\ee
yields 
\be
\lim\limits_{p\to0}G^{-1}_\alpha(p)=\tml(\alpha)+Z(\alpha)p^2+O(p^4),
\ee
for the fluctuations in ${\cal B}_1$ and ${\cal B}_{16}$. The mass and the wave function 
renormalization constant are given in Table 1 for $\alpha=1$ and 16. 

Notice that the finiteness of the mass in ${\cal B}_{16}$ requires a
tree-level renormalization of $c_2$, such that $\cm(4,c_2)=O(a^2)$. The continuum
limit of the mean-field solution is achieved at the critical 
point CR of Fig. 5.

In the other restricted Brillouin zones in each phase we get
\bea
\lim\limits_{p\to0}G^{-1}(P(\alpha)+p)
&=&p^2[1-8c_2d(\alpha)]\nonu
&&-p'^2[1-8c_2d(\alpha)]\nonu
&&+4d(\alpha)-16c_2d^2(\alpha)+\tilde M^2(\alpha),
\eea
where $d(\alpha)$ is given by (\ref{ncomps}) and $p'=P(16)-p$.
The particular form of $\tilde M^2(\alpha)$ depends on the phase 
and diverges as $O(a^{-2})$
when the masses in the $\alpha=1$ and $\alpha=16$ regions are kept 
finite. 

\underline{The mass spectrum:} We define the chiral lines  
$\chi_P$, $\chi_F$ and $\chi_{AF}$ as the lines where the masses of the 
two particles $\tml(1)$ and $\tml(16)$ are degenerate in each of the three phases 
considered above. These lines are actually given by the equation $\cm=0$
(see Fig.5b). As this line for $\ml<0$ is also the F-AF phase transition line, 
$\chi_F$ and $\chi_{AF}$ are actually one and the same line. 
The energy density and the particle content approaching this line from the two
sides are the same. It is worth to remind that in this case the even and odd 
sublattices are decoupled and that the difference between the upper and lower side
of this line lies on the sign of the ferromagnetic condensate on each of these 
sublattices.

The particle of the restricted zone
${\cal B}_1$ is the lighter one in the ferromagnetic phase and on the left of the 
chiral line $\chi_P$ in the paramagnetic phase. The staggered 
excitations of ${\cal B}_{2^d}$ are the lighter ones in the antiferromagnetic phase 
and on the
right of $\chi_P$. The
excitations of the restricted zone ${\cal B}_{2^d}$ are always massless along the
transition line $P-AF$.

\subsection{The momentum conservation}
The momentum is not conserved in the antiferromagnetic phase because the particles
may exchange momentum with the inhomogeneous vacuum. One can recover the 
momentum conservation by the introduction of the momentum
\be
p_\mu\longrightarrow p_{AF\mu}=p_\mu ({\rm mod}\pi),
\ee
where the quanta of the momentum which can be borrowed from the vacuum
is removed. Whenever this happens the particle type changes. The simultaneous shift
of all components, $p\to p+P(2^d)$, corresponds to the exchange of the two particles,
$1\longleftrightarrow2^d$.

\section{The symmetries}
The phase structure and the order parameter of the model is quite
involved so it is all the more important to find the symmetries 
relevant to the phase transitions. We can identify two kind of symmetries,
one which is realised at certain points only of the phase boundary 
and others which distinguish the different phases.

\subsection{Chiral symmetry}
There are two particles in the model so one expects that the theory where 
the two particle species become symmetrical might be special. The transformation
\be
\chi:~~~~~~~~\phi(x)\longrightarrow(-1)^{\sum_\mu x^\mu}\phi(x),\label{chtrrs}
\ee
which amounts to the shift
\be
p_\mu\to p_\mu+P_\mu(2^d)
\ee
connecting the particle species will be called chiral transformation \cite{chiral}. 
It always leaves the ultralocal even potential energy invariant. The propagator and 
with it the kinetic energy changes as
\bea
G^{-1}(p)&=&{\cal P}^2(p)-c_2{\cal P}^4(p)+\tml\nonu
&\to&1-{\cal P}^2(p)-c_2[1-{\cal P}^2(p)]^2+\tml\nonu
&=&(-1+8dc_2){\cal P}^2(p)+(-c_2){\cal P}^4(p)+\cm(d,c_2)+\tml.
\eea
The theory which is invariant under $\chi$,
\be
c_2={1\over4d},~~~~c_4=0,\label{chinth}
\ee
will be called chiral symmetrical. Note that mass parameter of the kinetic 
energy is vanishing, $\cm(d,c_2)=0$ and the two particle species are
degenerate in such a theory.

The operator ${\cal P}_\pm=\hf(1\pm\chi)$
projects on the fields belonging to the even or odd sublattices, 
\be
{\cal P}_\pm\phi_\pm=\phi_\pm.\label{slproj}
\ee
The kinetic energy couples $\phi_+$ and 
$\phi_-$ in general. Since the transformation (\ref{chtrrs}) acts as 
\be
\phi_\pm\to\pm\phi_\pm,
\ee
the fields $\phi_+$ and $\phi_-$ decouple in the chiral invariant theory.
This decoupling gives another insight into the dynamics of the phase transitions.
The chiral theory contains two independent and equivalent $\phi^4$ theories. 
If they are in the symmetry broken phase then their condensate has the same
absolute magnitude. The relative phase is undetermined and will be the
result of the microscopic differences between the fluctuations of the two fields, in 
a manner similar to a spontaneous symmetry breaking. The ferromagnetic phase
is realised when the sign of the condensates agree. The sign is the opposite
in the antiferromagnetic case. The spontaneous symmetry breaking is the
result of the infrared modes in the independent theories. In case
when the sign of the condensate happens to be different then the resulting
vacuum of the original theory which contains both sublattices has 
an ultraviolet condensate. In this manner the original, infrared mechanism
appears in the ultraviolet and generates dynamical symmetry breaking
for the observables of the complete lattice.

\subsection{Chiral bosons}
The origin of the chiral symmetry becomes clearer by introducing the 
hyper-cube variables $x^\mu=2y^\mu+n^\mu$ where $n^\mu$ labels the different
sites of the elementary cell of the $(1,2)$ antiferromagnetic vacuum 
and the chiral fields\cite{morel} 
\be
\phi_n(y)=\phi_\alpha(y)=\phi(2y+n(\alpha)).
\ee
We need the linear superpositions \cite{thun}
\be
\tilde\phi_\alpha(y)=A_{\alpha\beta}\phi_\beta(y),
\ee
where the matrix
\be
A_{\alpha\beta}=2^{-d/2}(-1)^{n(\alpha)\cdot n(\beta)}
\ee
performs the $Z_2$ Fourier transformation in the elementary cell. Since 
\be
(A^2)_{n,n'}=2^{-d}\sum\limits_m(-1)^{m\cdot(n+n')}=\delta_{n,n'},
\ee
the inverse Fourier transformation is
\be
\phi_\alpha(y)=A_{\alpha\beta}\tilde\phi_\beta(y).
\ee

The chiral transformation is diagonal on the chiral field basis,
\be
\chi:~~~~~~~~\phi_\alpha(y)\longrightarrow\chi(n(\alpha))\phi_\alpha(y),\label{khitra}
\ee
where
\be
\chi(n)=(-1)^{\sum_\mu n^\mu}=(-1)^{E(0)\cdot n}.\label{khikv}
\ee
The vector of the last expression is defined as 
\be
E_\mu(k)=\cases{1&$k\le\mu$,\cr0&otherwise.}
\ee

\subsection{Space-time inversions}
The space-time inversions will serve two purposes: On the one hand, they
demonstrate the formal similarity between the field $\phi_\alpha(y)$
and the chiral fermions. On the other hand, they are the symmetries which
distinguish the ferromagnetic and the antiferromagnetic phase.
The inversion $I_\nu$ of the coordinate 
\be
I_\nu:~~~~~~~~x^\mu\longrightarrow I_\nu x^\mu=(-1)^{\delta_{\mu\nu}}x^\mu,
\ee
is defined in such a manner that it maps the elementary cells
into each other. It
flips the $\mu$-th components of the elementary cell vector $n_\mu$, 
\be
I_\mu:~~~~~~~~\phi(y)\longrightarrow U_\mu\phi(I_\mu y),
\ee
where the matrix $U_\mu$ acting on the elementary cell is defined as
\be
(U_\mu)_{n,m}=\cases{1&if $n_\nu+m_\nu=\delta_{\mu,\nu}\mod$,\cr
0&otherwise.}
\ee
The space inversion, $P=\prod\limits_{\ell=2}^dI_\ell$, is
\be
P:~~~~~~~~\phi(y)\longrightarrow U_P\phi(Py),
\ee
where 
\be
(U_P)_{n,m}=(\prod\limits_{\ell=2}^d U_\ell)_{n,m}=\cases{1&if $n+m
=E(1)\mod$,\cr0&otherwise.}
\ee
The field $U_P\phi(y)$ will be called the P-helicity partner of $\phi(y)$. 
The combined effect of the time inversion $T=I_1$ and the space inversion
is represented by
\be
PT:~~~~~~~~\phi(y)\longrightarrow U_{PT}\phi(PTy),
\ee
\be
(U_{PT})_{n,m}=(\prod\limits_\mu U_\mu)_{n,m}=\cases{1&if $n+m
=E(0)\mod$,\cr0&otherwise.}
\ee
$U_{PT}\phi(y)$ will be called the PT-helicity partner of $\phi(y)$. 
Finally we define $\rho$ as 
\be
\rho=\cases{P& even d,\cr PT& odd d,}
\ee
which maps the fields of the two sublattices, $\phi_+$ and 
$\phi_-$ into each other in a specific manner.

We found that the chiral transformation is diagonal and the space-time inversions
are non-diagonal in the chiral field basis. The situation is just the opposite after a 
Fourier transformation. First we show that the Fourier transformed fields are 
eigenvectors of the space-time inversions. $\tilde U_\mu$ which 
represents $I_\mu$ on the Fourier transformed elementary cell is given by 
$\tilde U_\mu A=AU_\mu$, what yields 
\bea
(\tilde U_P)_{n,n'}&=&(AU_PA)_{n,n'}\nonu
&=&2^{-d}\sum\limits_m(-1)^{m\cdot(n+n')+n\cdot E(1)}\nonu
&=&\delta_{n,n'}(-1)^{n\cdot E(1)}\nonu
&=&\delta_{n,n'}(-1)^{\sum\limits_{\ell=2}^d n_\ell}.
\eea
In a similar manner we have
\be
(\tilde U_{PT})_{n,n'}=\delta_{n,n'}(-1)^{\sum\limits_{\mu=1}^d n_\mu}
=\delta_{n,n'}\chi(n).
\ee
Thus the Fourier transformed fields have well defined space and time 
inversion parities, 
\be
\sigma_P=(-1)^{\sum\limits_{\ell=2}^d n_\ell},~~~~~~~~\sigma_T=(-1)^{n^1}.
\ee

On the contrary, the chiral transformation 
becomes non-diagonal after a Fourier transformation,
\be
\chi:~~~~~~~~\tilde\phi_\alpha(y)\longrightarrow\tilde\phi_{\bar\alpha}(y)
\ee
with $\bar\alpha=2^d+1-\alpha$. The corresponding transformation of the 
vector index $n_\mu(\alpha)$ is
\be
\chi:~~~~~~~~n\longrightarrow\bar n=n+E(0)\mod.
\ee

\subsection{Bosonic chiral theory}
It is worthwhile to compare our result with the fermionic case. 

{\em Particle species:} The naive
fermion theory has $2^d$ species in lattice regularization which is just the number
of restricted Brillouin zones in the antiferromagnetic phase $(1,2)$.
Out of the $2^d$ restricted Brillouin zones only two contain particle modes for
small $c_4$. 
The helicity, the projection of the angular momentum on the momentum of a scalar particle
is identically vanishing. Nevertheless one can construct a pair of scalar fields 
$\phi_\pm(x)=\phi_s(x)\pm\phi_{ps}(x)$, which are exchanged
under space inversion by the help of a scalar and a pseudoscalar field, $\phi_s(x)$,
$\phi_{ps}(x)$, respectively. The chiral spinors are exchanged by the space inversion, $U_P=i\gamma_0$. 
By analogy we may call $\phi_\pm(x)$ and $\phi_\alpha(y)$
chiral fields. The excitations around $p=P(1)$ and $P(16)$ of the Brillouin zone
correspond to the slowly varying fields $\phi_s(x)$ and $\phi_{ps}(x)$, respectively.
Note that the projection of the angular momentum on the momentum is the same, 0,
for all members of the $O(d)$ multiplet of the chiral scalar particle as for the 1+1 dimensional
fermions. 

{\em Chiral symmetry:}
The chiral spinors decouple in the massless fermionic theory as our chiral boson
fields do at the theory (\ref{chinth}). The analogue of the discrete chiral
transformation, $\psi\to\gamma_5\psi$, is $\phi\to\chi\phi$, given by
(\ref{chtrrs}). The standard representation for the Dirac bispinors
provides fields with well defined parity similar to the Fourier transformed
chiral fields of the scalar model. 

{\em Chiral charge:} The chiral charge of a chiral spinor is its eigenvalue for $\gamma_5$.
The sum of the chiral charge is zero for the naive fermions on the lattice.
The chiral charge of the scalar modes on the even or the odd sublattices is +1 or -1, 
respectively. Thus the total chiral charge is vanishing in our model as long as 
there are as many degrees of freedom on the even as on the odd sublattice. 

{\em Chiral particles:}
The chiral fermions represent a serious problem in lattice regularization because 
the theory with a single chiral fermion is not covariant under space inversion. The
realization of this condition meets difficulties in the usual lattice theories \cite{nogo}.
There is more flexibility in the scalar model where 
we might as well use the symmetrical or the anti-symmetrical combination of the two 
particle modes in ${\cal B}_1$ and ${\cal B}_{2^d}$ for the chiral symmetrical theory. In fact,
these modes are degenerate and decouple. Such a combinations correspond
to the fields $\phi_\pm$ constructed in (\ref{slproj}). Since these fields
have no interactions between themselves one of them can be set to zero.
The resulting model which exists only at the chiral point contains
a single particle with non-vanishing chiral charge and a freely adjustable mass
and coupling constant,
\be
\cL_\pm=\hf(\partial_\mu\phi_\pm)^2+{m^2\over2}\phi^2_\pm+{\lambda\over4}\phi^4_\pm,
\ee
in the continuum limit of the mean-field approximation. 

The no-go theorem, \cite{nogo},
about the impossibility of having a single fermion with non-vanishing
chiral charge has topological origin. The non-trivial elementary cell of the
antiferromagnetic phase which becomes small in the continuum limit offers the 
possibility of avoiding the usual transformation rules with respect the 
space-inversions and thereby circumvents the problem at least for bosons.
The differences of the physical properties of a chiral and an ordinary boson
with well defined parity will only be seen after coupling the chiral boson to 
other particles.

When the theory with broken chiral symmetry is considered then the chiral fields
$\phi_\pm$ are coupled. The low energy elementary excitations are made of the slowly
varying fields $\phi_\pm$ as
\be
\tilde\phi_\pm=\phi_+\pm\phi_-.
\ee
The $\rho$ parity of the field $\tilde\phi_\pm$ is $\pm1$ and it belongs to the zone
${\cal B}_1$ and ${\cal B}_{16}$. The effective theory obtained on the
tree level, (\ref{varpo}), is 
\be
L=\hf(\partial_\mu\tilde\phi_+)^2+\hf(\partial_\mu\tilde\phi_-)^2+
{m^2_+\over2}\tilde\phi_+^2+{m^2_-\over2}\tilde\phi_-^2
+{\lambda\over4}(\tilde\phi_+^4+\tilde\phi_-^4+6\tilde\phi_+^2\tilde\phi_-^2),
\ee
for low enough energy where the further influence of the higher order derivatives
on the propagator is negligible.

\subsection{The symmetries of the phase diagram}
For the more complete characterization of the symmetries of the
model we finally introduce the discrete analogue of the charge conjugation 
to our real field,
\be
\gamma:~~~~~~~~\phi(x)\to-\phi(x).
\ee
The symmetries of different regions of the coupling constant space 
$(\ml,c_2)$ is shown for $c_4=0$ in Table 2.
One can see that the ferromagnetic condensate is detectable
by the internal space order parameter only. The dynamical breakdown of the space
inversion symmetry in the vacuum is characteristic of the antiferromagnetic phase. 
In agreement with this remark the particle scattering off such a  vacuum may borrow
the momentum $P(2^d)$ and change its parity.

\section{Conclusion}
We showed an interplay between the symmetry breaking patterns in the internal and
the external space realized by the higher dimensional pieces of the kinetic 
energy term of the action of a $\phi^4$ theory. The strongly distance dependent 
interactions described
by these pieces generate non-trivial elementary cells in the vacuum and render
the dynamics of the system somewhat similar to Solid State Physics. 

A complex phase structure was found with a number of 
commensurate incommensurate transitions. 
Concerning the elementary excitations are concerned there are two 
particle modes in the vicinity of the antiferromagnetic $(1,2)$ phase, 
the analogues of the acoustic and optical phonons
of the solid states because they correspond to the in phase and the out of 
phase oscillations in the elementary cell. The emergence of the non-homogeneous
condensate of the phase $(1,2)$ reduces the translation invariance into 
translation by even number of the lattice spacing. Nevertheless no Goldstone
bosons appear. This is because the condensate is at the cut-off scale where the
continuous translation symmetry is broken by the regularization.

The space inversion exchanges the two particles of the theory. This 
opens the possibility of constructing chiral bosons on the lattice for 
such a choice of the coupling constants where these two particles decouple. 
We showed that the dynamical breaking of the
space inversion symmetry is characteristic to the formation of the non-trivial 
elementary cells of the antiferromagnetic phase.

We believe that the phenomena mentioned in the framework of the scalar $\phi^4$
model are generic and can be found in any bosonic model. A theoretical test of
such a model as a more realistic effective theory is the possibility
of removing the length scale of the elementary cell of the vacuum in order to 
suppress the non-unitary processes related to the creation of the lattice 
defects \cite{chiral} \cite{unit} \cite{jochen}.
The period length of the vacuum can be sent to zero in the one-loop approximation
\cite{next}. It remains to be seen if this result generalizes to higher loop
order. One should bear in mind that such a perturbative approach is reasonable 
at or above the upper critical
dimension. For $d<4$ one expects to find strongly coupled long range modes and the
vacuum is made homogeneous in a manner similar to spin fluids \cite{fradkin}.

\begin{figure}
	\vspace{0.5cm}
	\epsfxsize=11cm
	\epsfysize=5cm
	\centerline{\epsffile{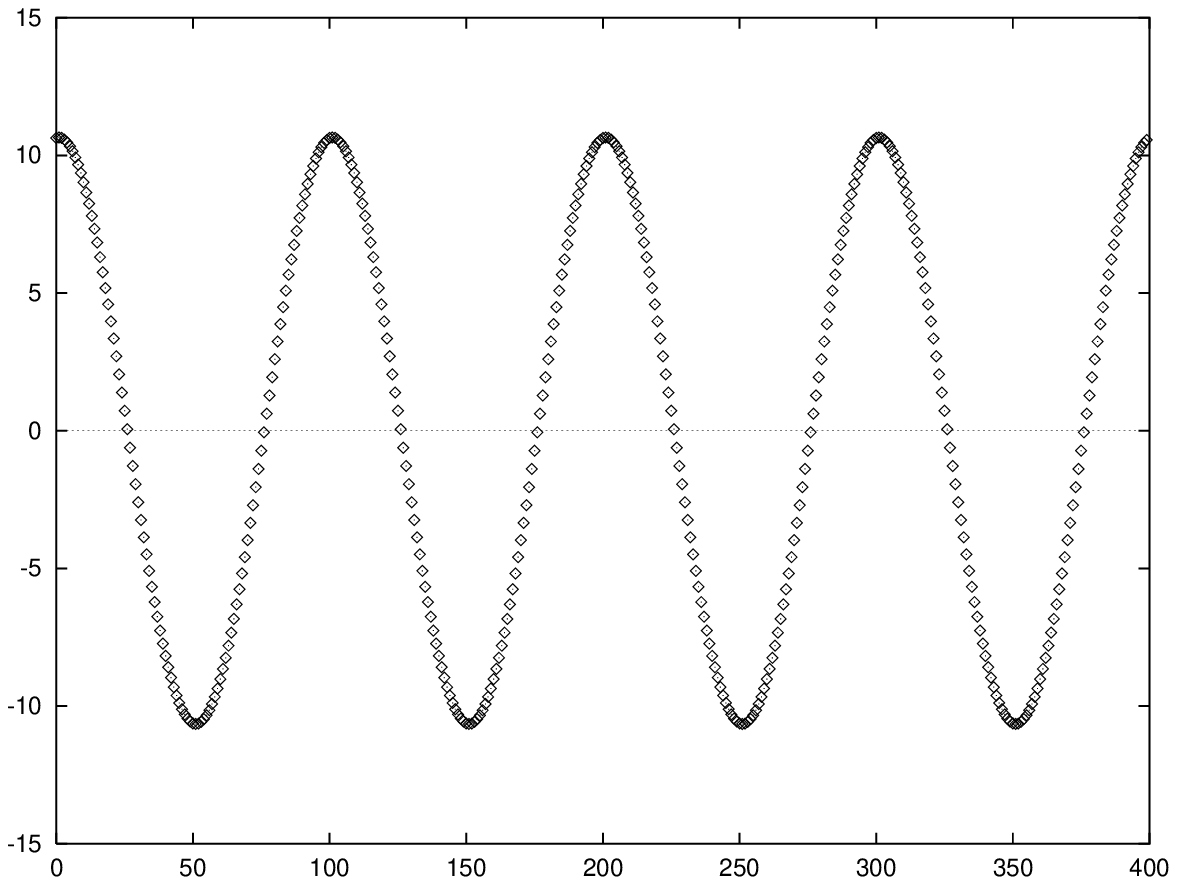}}
\centerline{(a)}
	\epsfxsize=11cm
	\epsfysize=3cm
	\centerline{\epsffile{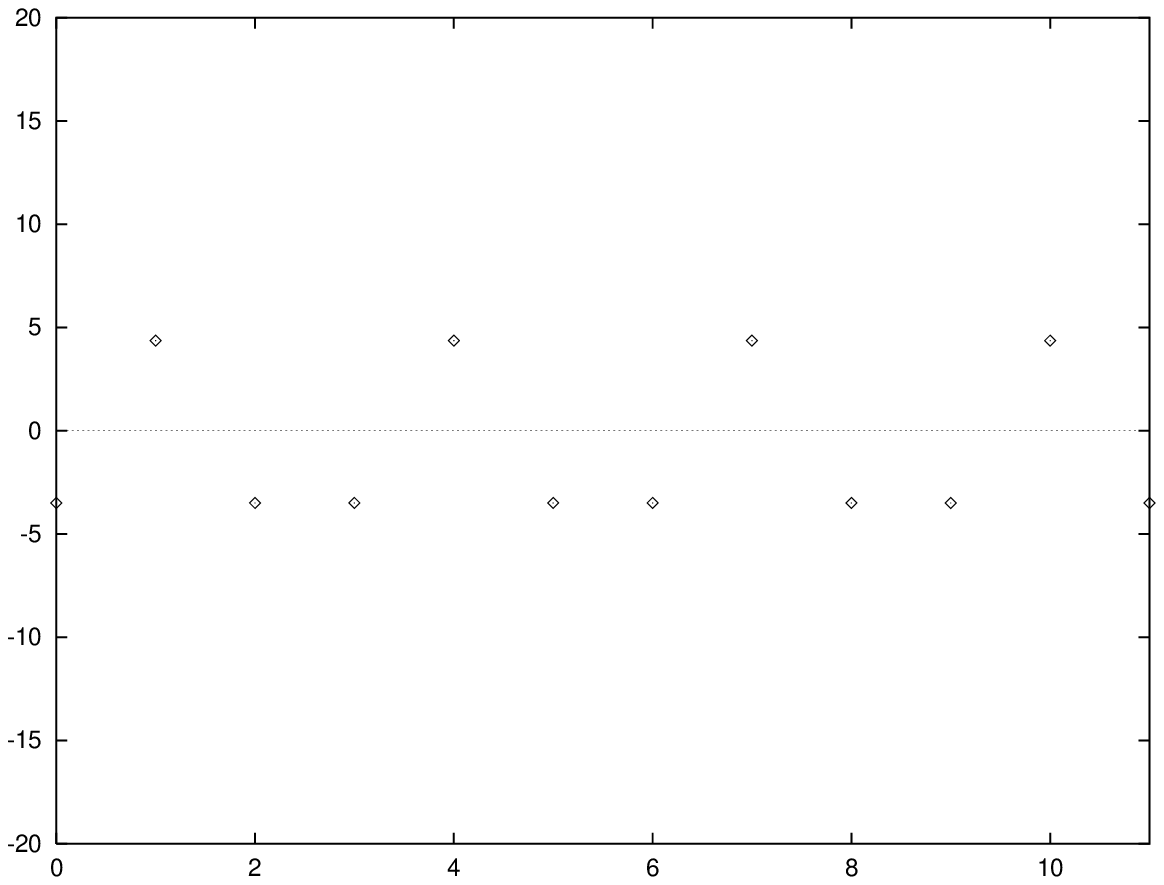}}
\centerline{(b)}
\caption{Field configuration in function of $x$ which minimizes of the action density 
with periodic boundary condition in d=1 dimension. (a): Continuum theory, (b): The
vacuum of the phase $(1,3)$ in lattice regularization.}
\end{figure}

\begin{figure}
	\vspace{0.5cm}
	\epsfxsize=5cm
	\epsfysize=6cm
	\centerline{\epsffile{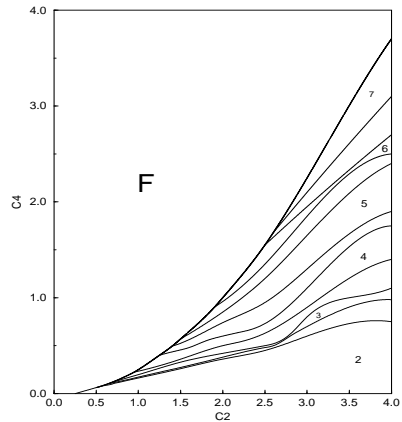}}
\caption{The phase diagram for the lattice regularized model in d=1.
F stands for the ferromagnetic phase and the numbers denote the parameter
$N$ of the antiferromagnetic phase $(1,N)$.}
\end{figure}
\begin{figure}
\begin{minipage}{6cm}
	\epsfxsize=5cm
	\epsfysize=6cm
	\centerline{\epsffile{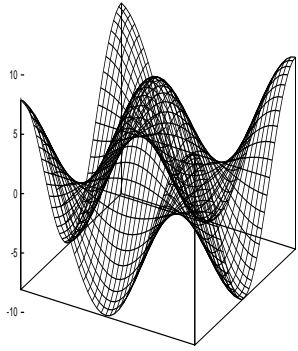}}
\centerline{(a)}
\end{minipage}
\hfill
\begin{minipage}{6cm}
	\epsfxsize=5cm
	\epsfysize=6cm
	\centerline{\epsffile{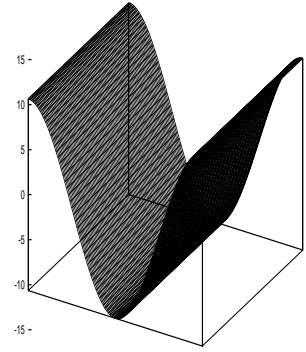}}
\centerline{(b)}
\end{minipage}
\vskip 20pt
\caption{The elementary cell of the antiferromagnetic vacuum configurations in the continuum
as the functions of $x^\mu$ in d=2. (a): Relativistic 
vacuum, a local minimum. (b): The non relativistic vacuum, the absolute minimum.}
\end{figure}

\begin{figure}
\begin{minipage}{4cm}
	\epsfxsize=4cm
	\epsfysize=4cm
	\centerline{\epsffile{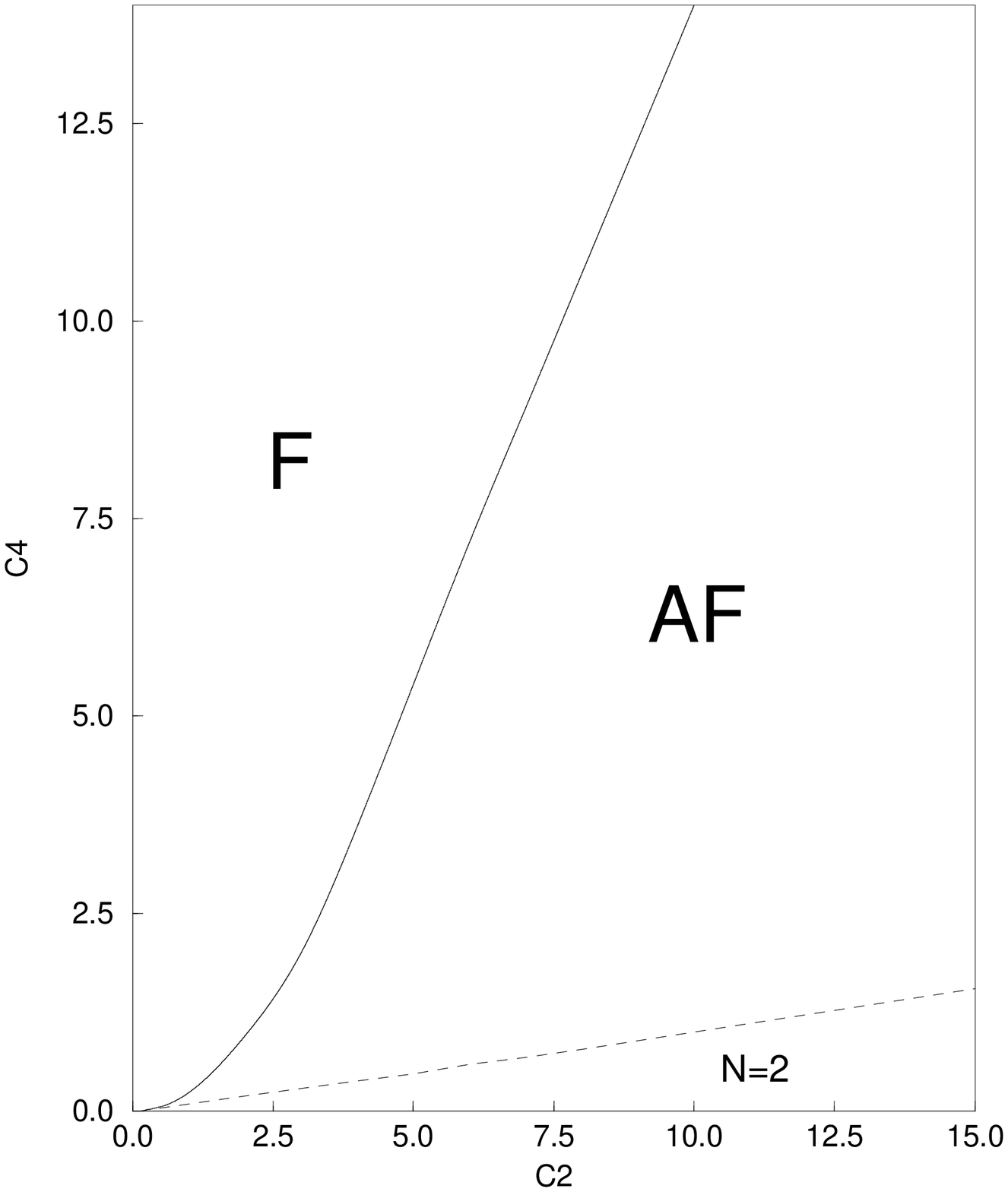}}
\centerline{(a)}
\end{minipage}
\hfill
\begin{minipage}{4cm}
	\epsfxsize=4cm
	\epsfysize=4cm
	\centerline{\epsffile{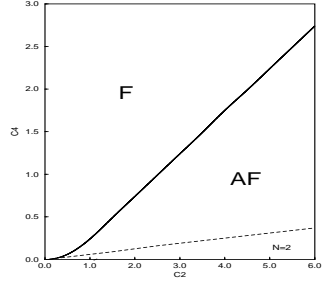}}
\centerline{(b)}
\end{minipage}
\hfill
\begin{minipage}{4cm}
	\epsfxsize=4cm
	\epsfysize=4cm
	\centerline{\epsffile{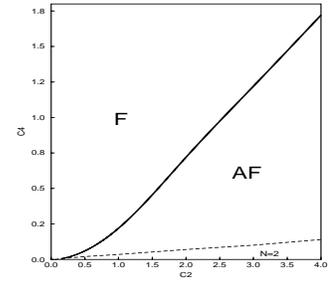}}
\centerline{(c)}
\end{minipage}
\vskip 20 pt
\caption{The phase diagrams in (a): d=2, (b): d=3, (c): d=4.
The antiferromagnetic phase is found below the solid line. The lower 
region below the dashed line is the phase $(1,2)$. The higher $N$ phases
are not shown.}
\end{figure}

\begin{figure}
\begin{minipage}{5.5cm}
	\epsfxsize=5cm
	\epsfysize=5cm
	\centerline{\epsffile{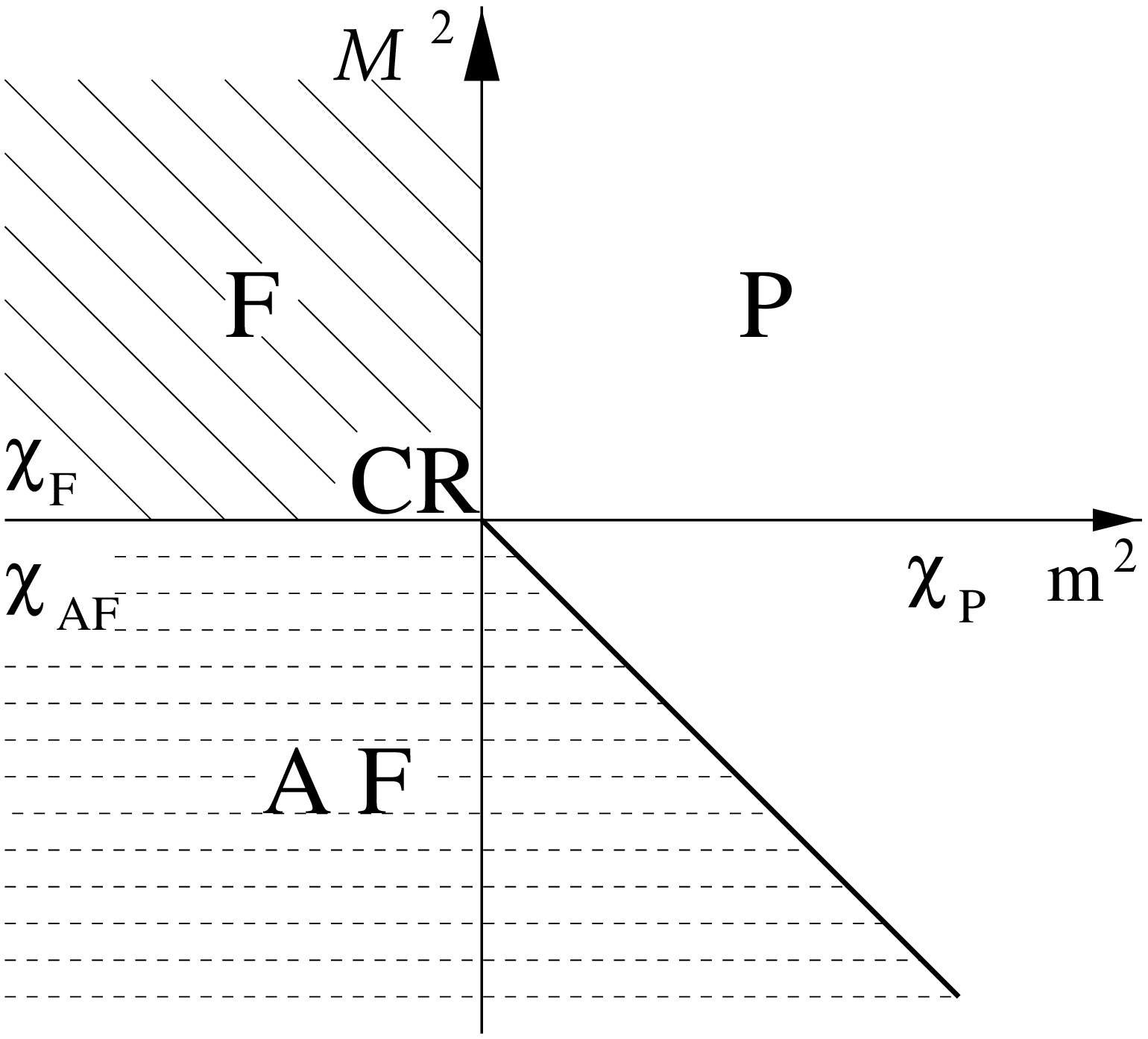}}
\centerline{(a)}
\end{minipage}
\hfill
\begin{minipage}{5.5cm}
	\epsfxsize=5cm
	\epsfysize=5cm
	\centerline{\epsffile{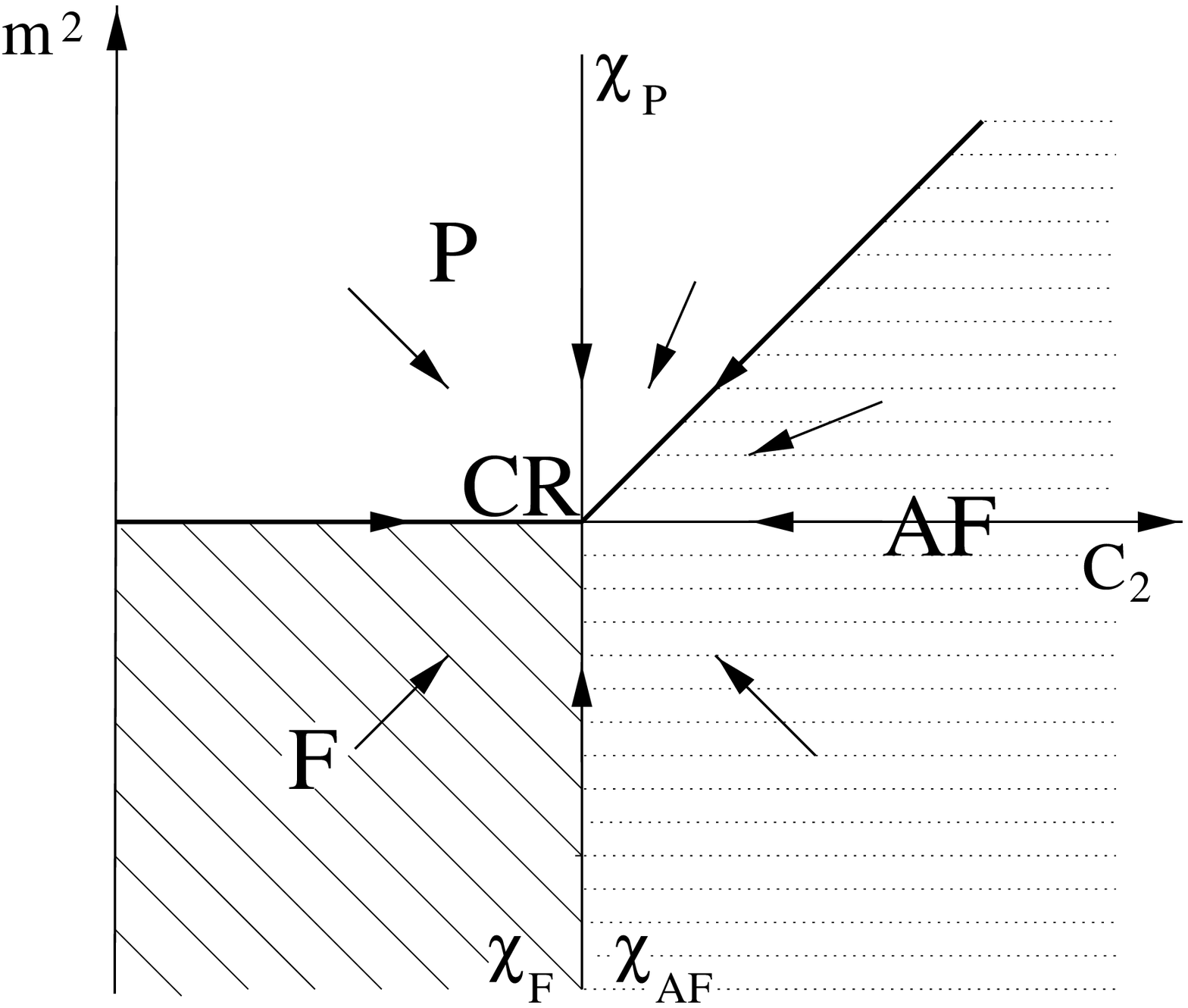}}
\centerline{(b)}
\end{minipage}
\vskip 20pt
\caption{The phase boundary between the paramagnetic (P), ferromagnetic (F) 
and the $N=2$ antiferromagnetic (AF) phase for $c_4=0$. The chiral symmetric 
regions $\chi_F$, $\chi_{AF}$ and the critical point CR are on the 
phase boundary. (a): The plane $(\ml,\cm)$; (b): The plane $(c_2,\ml)$. 
The chiral line $\chi_P$ splits the paramagnetic phase into two parts.
On the left of $\chi_P$ the particle of the restricted zone ${\cal B}_{1}$ is 
the lighter one. The particle of the zone ${\cal B}_{2^d}$ is the lighter one on the 
other side. The arrows show the 
possible continuum limits at the chiral invariant critical point, CR.}
\end{figure}

\begin{figure}
\begin{minipage}{4cm}
	\epsfxsize=4cm
	\epsfysize=4cm
	\centerline{\epsffile{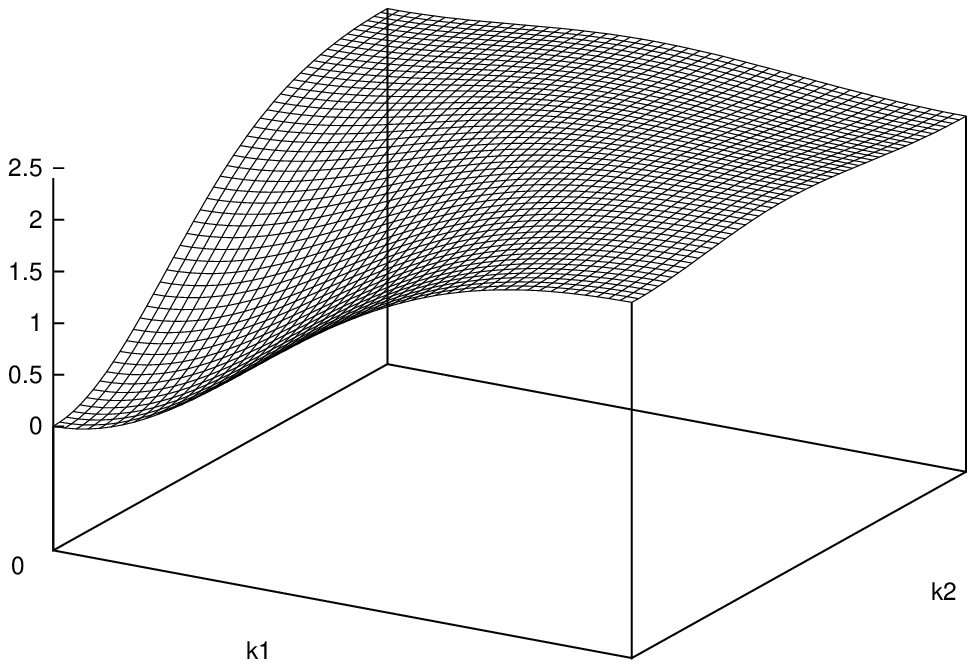}}
\centerline{(a)}
\end{minipage}
\hfill
\begin{minipage}{4cm}
	\epsfxsize=4cm
	\epsfysize=4cm
	\centerline{\epsffile{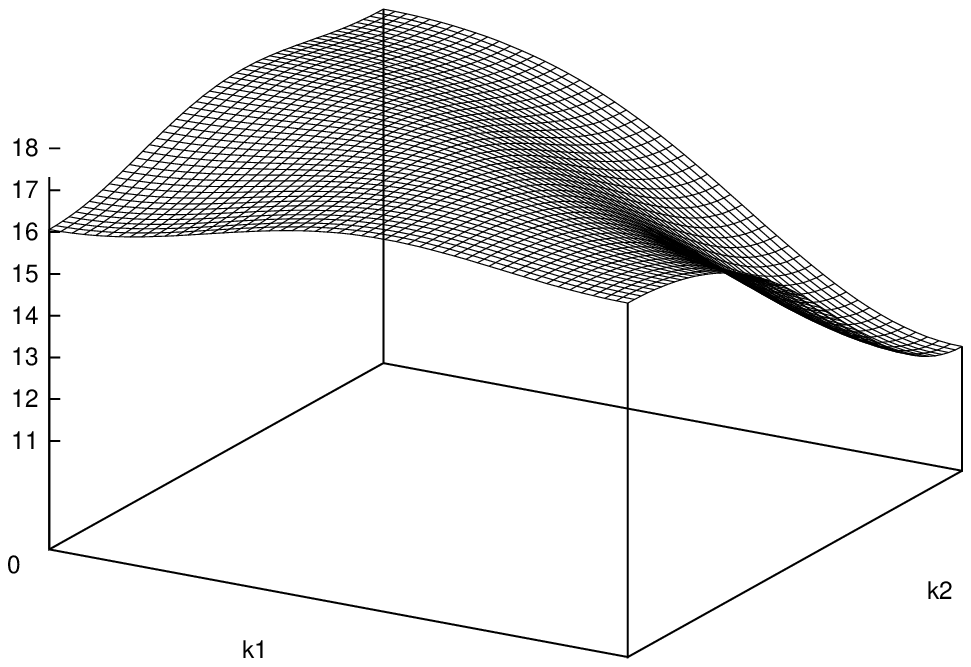}}
\centerline{(b)}
\end{minipage}
\hfill
\begin{minipage}{4cm}
	\epsfxsize=4cm
	\epsfysize=4cm
	\centerline{\epsffile{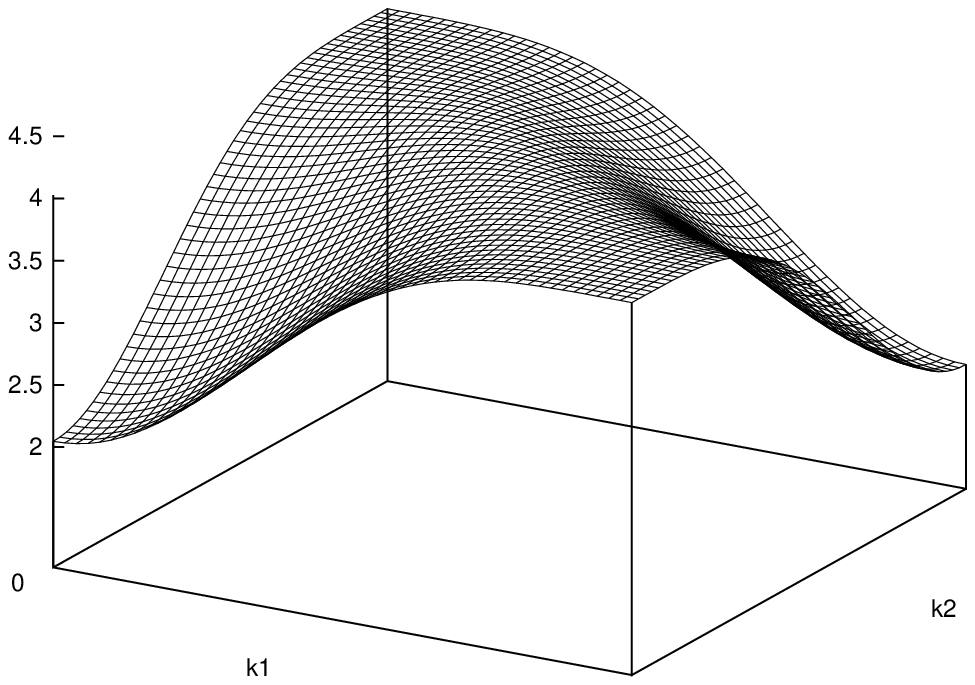}}
\centerline{(c)}
\end{minipage}
\vskip 20pt
\caption{The propagator in d=2 dimensions in the (a): ferro-, (b): antiferromagnetic
phase and (c): at the phase boundary.}
\end{figure}

\begin{table}
\begin{center}
\begin{tabular}{|c||c|c|c|c|}
%\begin{tabular}{@{}*{5}{|l||l|l|l}}
\hline    
phase&$\tml(1)$&$\tml(16)$&$Z(1)$&$Z(16)$\cr 
%\hline   
\hline
P &$\ml$       &$\ml+\cm$   &1&$-1+32c_2$\cr
F &$-2\ml$     &$-2\ml+\cm$ &1&$-1+32c_2$\cr
AF&$-2\ml-3\cm$&$-2\ml-2\cm$&1&$-1+32c_2$\cr
\hline
\end{tabular}
\vskip 30pt
\caption{The parameters of the propagator for ${\cal B}_\alpha$, $\alpha=1$ 
and $16$.}
\end{center}
\end{table}

\begin{table}
\begin{center}
\begin{tabular}{|c||c|c|c|}
%\begin{tabular}{@{}*{4}{|l||c|c|c}}
\hline    
phase&$~~~~\chi~~~~$&$~~~~\rho~~~~$&$~~~~\gamma$~~~~\cr 
%\hline    
\hline
P          &E      &$\surd$&$\surd$\cr
F          &E      &$\surd$&S      \cr
AF         &E      &D      &D      \cr
CR         &$\surd$&$\surd$&$\surd$\cr
$\chi_P$   &$\surd$&$\surd$&$\surd$\cr
$\chi_F$   &$\surd$&$\surd$&S      \cr
$\chi_{AF}$&$\surd$&D      &D      \cr
\hline
\end{tabular}
\vskip 30pt
\caption{The symmetry of different regions of the coupling constant
space. A symmetry can be manifest ($\surd$), broken 
explicitly (E), spontaneously by the IR modes (S), or dynamically
by the UV modes (D).}
\end{center}
\end{table}

\acknowledgments
J.P. thanks Jochen and Werner Fingberg for illuminating discussions.

\end{document}